\documentclass{elsart}
\usepackage{graphicx}

\newcommand{\Meeg }{M_{ee \gamma}}
\newcommand{\Mppeeg }{M_{\pi\pi ee \gamma}}
\newcommand{\Mppee  }{M_{\pi\pi ee }}
\newcommand{\MKL }{M_{K_L}}
\newcommand{\Mpiz}{M_{\pi^0}}
\begin{document}

\begin{frontmatter}
\journal{Physics Letters B}
\title{Observation of the decay mode 
$K_L \rightarrow \pi^+ \pi^- e^+ e^-$ }
\author{Y. Takeuchi},
\author{Y. Hemmi\thanksref{DIT}},
\author{H. Kurashige\thanksref{KOBE}},
\author{Y. Matono},
\author{K. Murakami},
\author{T. Nomura\thanksref{CONTACT}},
\author{H. Sakamoto}, 
\author{N. Sasao},
\author{M. Suehiro}
\address
 {Department of Physics, Kyoto University, Kyoto 606-8502, Japan }
\author{Y. Fukushima}, 
\author{Y. Ikegami},
\author{T. T. Nakamura}, 
\author{T. Taniguchi}
\address
 {High Energy Accelerator Research Organization (KEK),
 Ibaraki 305-0801, Japan }
\author{M. Asai}
\address
 {Hiroshima Institute of Technology, Hiroshima 731-5193, Japan }
\thanks[DIT]
 {Present address: {\it Daido Institute of Technology, Aichi 457, Japan}}
\thanks[KOBE]
 {Present address: {\it Kobe University, Hyogo 657-8501, Japan}}
\thanks[CONTACT]
 {Contact person: nomurat@scphys.kyoto-u.ac.jp}
\begin{abstract}
We report on results of an experimental search for 
 the $K_L \rightarrow \pi^+ \pi^- e^+ e^-$ decay mode.
We found $13.5 \pm 4.0$ events and determined its branching ratio
 to be
 $ (4.4  \pm 1.3(\mbox{stat.}) \pm 0.5(\mbox{syst.}) ) \times 10^{-7}$.
The result agrees well with the theoretical prediction.
\end{abstract}
\end{frontmatter}

\section{Introduction}
In the previous Letter~\cite{Nomura97}, we reported the results 
 of our experimental search for the decay mode
 $K_L \rightarrow \pi^+ \pi^- e^+ e^-$,
 which established the upper limit of
 $4.6 \times 10^{-7}$ (90\% CL) on its branching ratio.
In this decay mode, CP violation may occur as an interference 
 between two intermediate states with different 
 CP properties~\cite{SW92,OtherTh} .
Thus it can provide a new testing ground for
 investigating CP violation.
A theoretical model predicts a branching ratio of 
 about $3 \times 10^{-7}$~\cite{SW92}.
Encouraged by this value, we continued the experiment
 further aiming to establish this decay mode.
Recently a group at the Fermilab (KTeV) 
 reported the measurement of its branching ratio based on  
 $36.6 \pm 6.8$ observed events~\cite{Adams98}.
In this article, we present our new result~\cite{Sasao97}.

\begin{figure*}[htbp]
 \begin{center}
  \includegraphics[scale=0.8, clip]{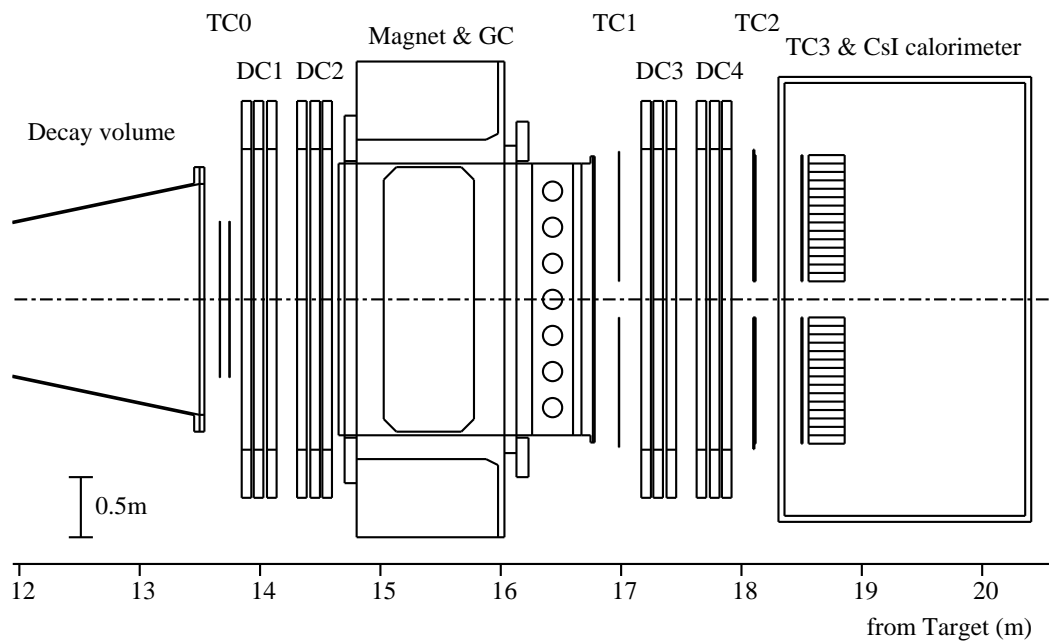}
  \caption{Schematic plan view of the KEK-E162 detector}
  \label{fig-setup}
 \end{center}
\end{figure*}

The experiment was conducted with a 12-GeV proton synchrotron 
 at High Energy Accelerator Research Organization (KEK).
The experimental set-up, shown in Fig.~\ref{fig-setup},
 was already described in Ref.~\cite{Nomura97}.
We used the same set-up with one vital change:
 the decay volume, 
 which had been filled with helium gas in the preceding run, 
 was evacuated down to $8 \times 10^{-3}$ Torr.
As will be seen, this was very effective to reduce backgrounds originating
 from nuclear interactions.
Below we briefly describe our set-up for convenience.
The $K_L$ beam was produced by focusing 12-GeV/c 
 primary protons onto a 60-mm-long copper target. 
The divergence of neutral beam was
 $\pm$4~mrad horizontally and $\pm$20~mrad vertically.
Following the 4-m-long decay volume was a charged particle spectrometer.
It consisted of four sets of drift chambers (DC1-DC4) and 
 an analyzing magnet with an average horizontal
 momentum kick of 136~MeV/c.
The drift chamber position resolution was in average 210~$\mu$m,
 and its efficiency was  greater than 98\% under a
 typical running condition. 
A threshold Cherenkov counter (GC) with pure N$_2$ gas at 1 atm 
 was placed inside of the magnet gap to identify electrons.
For the present decay mode, we obtained,
 by adjusting software cuts in the off-line analysis,
 an average electron efficiency of 99.7\% with 
 a pion-rejection factor of 50.
Two banks of pure CsI electromagnetic calorimeters, located at the far end,
 measured energy and position of electrons and photons.
Its energy and position resolutions were found to be 
 approximately 3\% and 7~mm for 1-GeV electrons, respectively.
There were 4 sets of trigger scintillation counters, 
 called TC0X,TC1X,TC2X/2Y and TC3X, 
 where X(Y) represented a horizontally(vertically)-segmented hodoscope.
The trigger for the present mode was designed to select
 events with at least three charged tracks which included at least two
 electrons.
\section{General Analysis and Normalization Mode}
In order to exploit full statistical power of the available data,
 we combined the present data (vacuum data) with the one 
 presented in Ref.~\cite{Nomura97} (He data), 
 and analyzed them without distinction.
Approximate $K_{L}$ flux ratio of the vacuum to He data set was $5:1$.
We closely followed the off-line analysis procedure described in 
 Ref.~\cite{Nomura97}
 with refinements at several places.
At first, events were required to have 4 (and only 4) tracks 
 with a common vertex in the beam region inside the decay volume.
Then particle species were determined.
A pion was identified as a track which could project onto
 a cluster in the calorimeter (a matched track) with ${\rm E/p} < 0.7$, 
 where E and p were an energy deposit measured by the calorimeter
 and a momentum of the track determined by the spectrometer,
 respectively. 
An electron was identified as a matched track with E $\geq$ 200~MeV,
 $0.9 \leq {\rm E/p} \leq 1.1$, and GC hits in the corresponding
 cells. 
The events containing $\pi^+$, $\pi^-$, $e^+$ and $e^-$ tracks
 were selected, 
 and then applied two kinematical cuts.
One was the requirement for the invariant mass of $e^+ e^-$ pair ($M_{ee}$) 
 to be at least 4~MeV$/c^2$; this cut was effective
 to suppress backgrounds from external conversion of $\gamma$-rays 
 into $e^+ e^-$ pair.
The other was a limit on the pion momentum asymmetry
($A_{+-} \equiv (p_{{\pi}^+}-p_{{\pi}^-})/(p_{{\pi}^+} + p_{{\pi}^-})$)
 within $\pm 0.5$. 
This  cut was introduced in Ref.~\cite{Nomura97}
 to suppress backgrounds due to nuclear interactions,
 but was found also useful to remove events with
 pion decays.  
The remaining events are our basic event samples for further analysis,
 and mainly contain both signal mode 
 $K_L \rightarrow \pi^+ \pi^- e^+ e^-$ and normalization mode 
 $K_L \rightarrow \pi^+ \pi^- \pi^0$
 ($\pi^{0} \rightarrow  e^+ e^- \gamma$).
\begin{figure*}[btph]
 \begin{center}
  \includegraphics[scale=1.0, bb=1 0 258 238, clip]{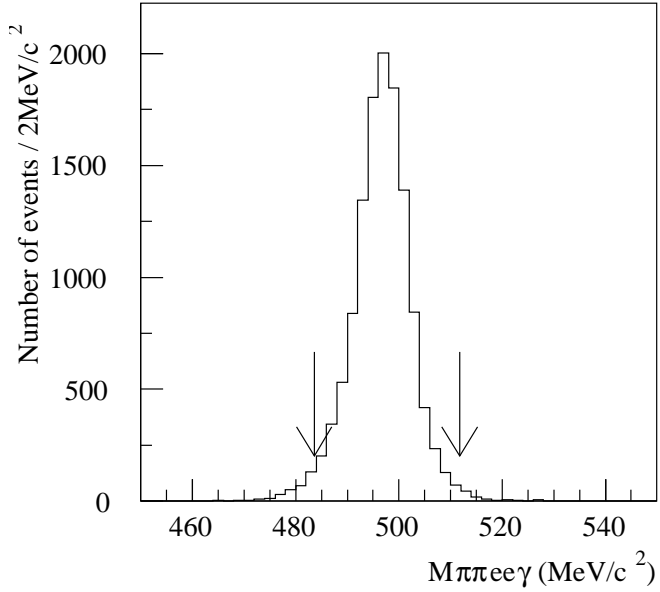}
  \caption{ The invariant mass distribution of $\pi^+ \pi^- e^+e^-\gamma$.
 The arrows indicate the region to select the 
  $K_L \rightarrow \pi^+ \pi^- \pi^0_D$ events.}
  \label{fig-ppeeg-mass}
 \end{center}
\end{figure*}

We now identify the normalization mode 
 (denoted as  $\pi^+ \pi^- \pi^0_D$ in the following).
We required at least one $\gamma$-ray with energy above 200 MeV.
Here a $\gamma$-ray was defined as 
 a cluster in the calorimeter 
 which did not match with any charged tracks.
If more than one $\gamma$-ray existed, 
 we selected the one for which 
 the invariant mass of $e^+ e^- \gamma$ ($\Meeg$)
 was closest to the $\pi^0$ mass.
Then we imposed three major kinematical cuts to select
  $\pi^+ \pi^- \pi^0_D$ events.
First, the invariant mass of $e^+ e^- \gamma$ ($\Meeg$)
 was required to be within $3 \sigma_{\Mpiz}$
 of the $\pi^0$ mass ($\Mpiz$), 
 where  the $\pi^0$ mass resolution  $\sigma_{\Mpiz}$
 was measured to be 4.6~MeV/c$^2$.
Second, $\theta ^2$ was required to be less than 20~mrad$^2$,
 where $\theta $ denotes the angle
 of the reconstructed $K_L$ momentum with respect to the line
 connecting the production target and decay vertex.
Fig.~\ref{fig-ppeeg-mass} shows the invariant mass distribution of 
 $\pi^+ \pi^- e^+ e^- \gamma$ ($\Mppeeg$)
 for the events which satisfied all the cuts mentioned so far. 
A clear $K_L$ peak can be seen with a mass resolution ($\sigma_{\MKL}$)
 of 5.5~MeV/c$^2$.
As a final cut, events were requested to lie within
  $3 \sigma_{\MKL}$ of the $K_L$ mass ($\MKL$).
After all the cuts, 12212 events remained~\cite{note-He-Dalitz}.
We call these events the $\pi^+ \pi^- \pi^0_D$ reconstructed events.

\section{Signal Mode and Background Subtraction}
\begin{figure*}[tbph]
 \begin{minipage}[b]{0.5\linewidth}
  \begin{center}
   \includegraphics[scale=0.8, bb=1 1 258 238, clip]{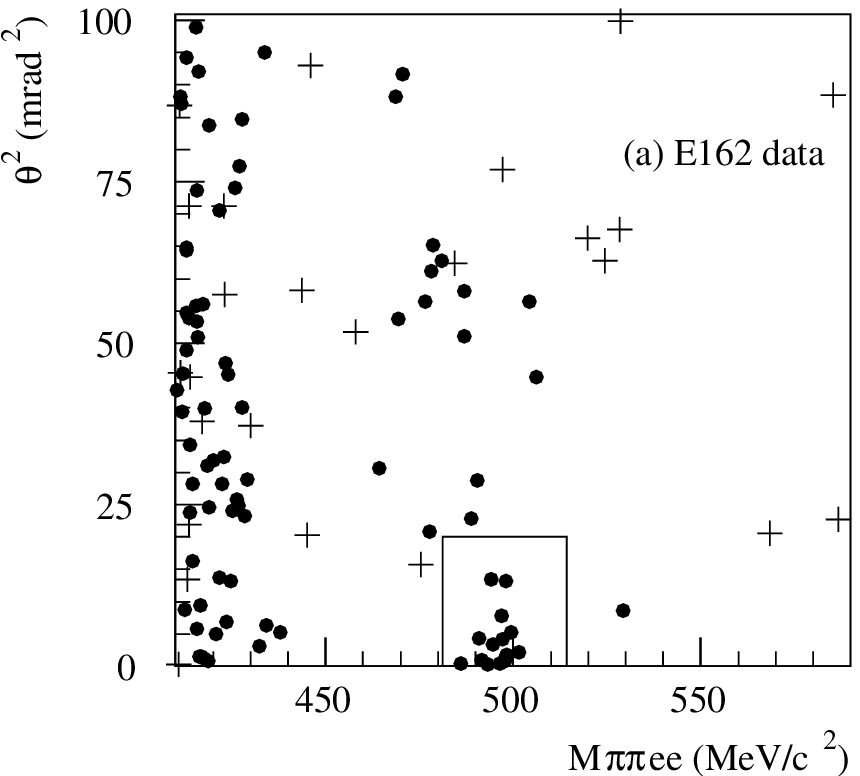}
  \end{center}
 \end{minipage}
 \begin{minipage}[b]{0.5\linewidth}
  \begin{center}
   \includegraphics[scale=0.8, bb=1 1 258 238, clip]{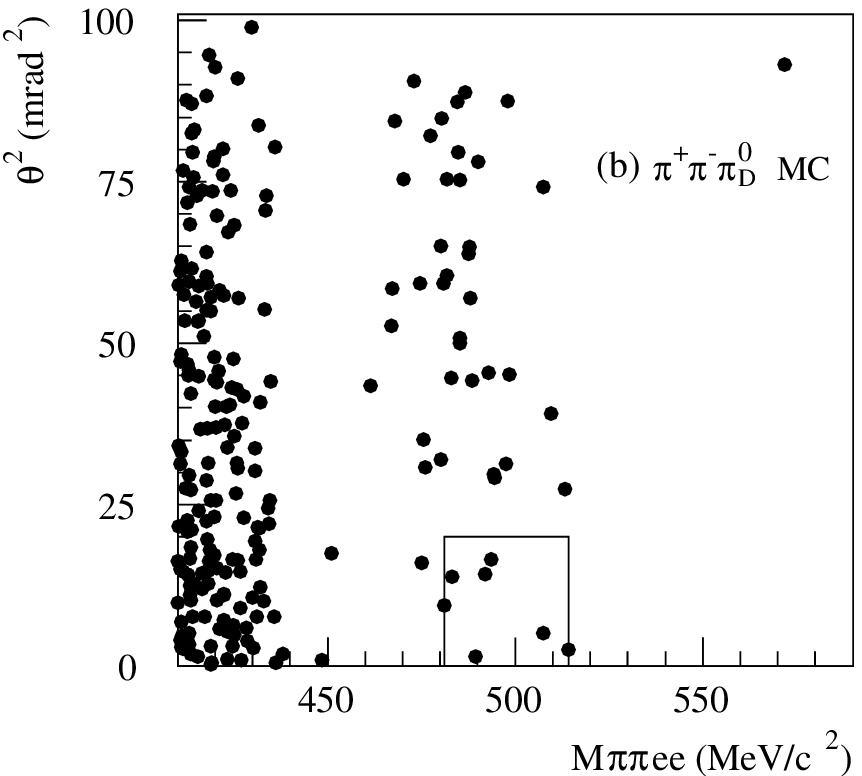}
  \end{center}
 \end{minipage}
 \caption{(a) The $M_{\pi\pi ee}$ vs $\theta^2$ scatter plot
 of the $K_L\rightarrow\pi^+\pi^-e^+e^-$ candidate
 events after the $\chi^2_D$ cut. 
 The solid dots represent the vacuum data and the plus the He data.
 The box indicates the signal region.
 (b) The corresponding scatter plot obtained by 
 the $\pi^+ \pi^- \pi^0_D$ MC simulation.
 Note that the MC statistics is 5 times more than the vacuum data.}
 \label{fig-ppee-scat}
\end{figure*}

For the signal mode $K_L \rightarrow \pi^+ \pi^- e^+ e^-$,
 the major backgrounds came from
 $\pi^+ \pi^- \pi^0_D$ events with an extra $\gamma$-ray
 missed from detection.
We employed the same strategy as in Ref.~\cite{Nomura97}.
At first, we removed events which had extra $\gamma$-ray(s)
 consistent with $\pi^0 \rightarrow e^+ e^- \gamma$.
To reduce the data set size, we also applied 
 at this stage loose kinematical cuts; 
 $\theta^2\leq$100~mrad$^2$ and $410<\Mppee<590$ MeV/c$^2$,
 where $\Mppee$ being the invariant mass of  $\pi^+ \pi^- e^+ e^-$.
Then we defined a parameter which quantified consistency of 
 an event with $\pi^+ \pi^- \pi^0_D$.
The parameter, called  $\chi^2_D$, was given by 
\[
  \chi^2_D(\vec{p}_{\gamma}) =
  \left( \frac{ \Meeg  -\Mpiz }{ \sigma_{\Mpiz} } \right)^2+
  \left( \frac{ \Mppeeg-\MKL  }{ \sigma_{\MKL } } \right)^2+
  \left( \frac{ \theta        }{ \sigma_{\theta} } \right)^2 ,
\] 
where $\vec{p}_{\gamma}$ was an arbitrary momentum 
 of a $\gamma$-ray assumed to exist~\cite{note-chi2D-change}.
The standard deviations $\sigma_{\Mpiz}$, 
 $\sigma_{\MKL }$ and $\sigma_{\theta}$~\cite{note-sigma-theta} were all
 obtained from the $\pi^+ \pi^- \pi^0_D$ reconstructed events.
We determined $\vec{p}_{\gamma}$ by minimizing $\chi^2_D$.
\begin{figure*}[htbp]
 \begin{center}
  \includegraphics[scale=1.0, bb=0 0 258 238, clip]{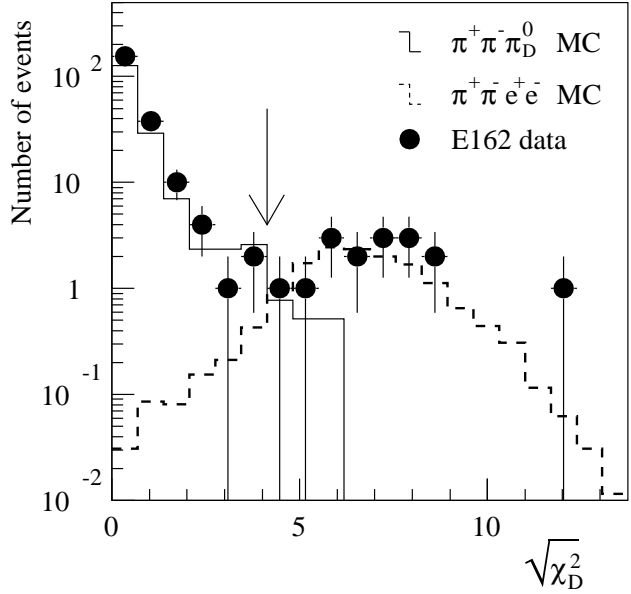}
  \caption{ The distribution of the square root of $\chi^2_D$.
  The solid and dashed lines are the results of Monte Carlo (MC) 
  simulation, respectively, for the $\pi^+ \pi^- \pi^0_D$ 
  and $\pi^+\pi^-e^+e^-$ modes. 
  The $K_L$ flux in the simulations was 
  normalized with the reconstructed $\pi^+ \pi^- \pi^0_D$ events,
  and the branching ratio for $\pi^+\pi^-e^+e^-$  
  was assumed to be given by our final result.
  The arrow shows the cut position.}
  \label{fig-chi2D}
 \end{center}
\end{figure*}
Fig.~\ref{fig-chi2D} shows the $\sqrt{\chi^2_D}$ distribution 
 of the events after minimization.\,
(Here only the events with $\theta^2\leq$20~mrad$^2$
 and $|\Mppee-\MKL|<5 \sigma_{\MKL }$ are plotted 
 to illustrate clearly existence of two components.)
The solid and dashed lines in the figure show 
 the results of Monte Carlo (MC)
 simulations, respectively, for the $\pi^+ \pi^- \pi^0_D$ 
 and $\pi^+\pi^-e^+e^-$ modes.
The $K_L$ flux in the simulations was 
 normalized with the reconstructed $\pi^+ \pi^- \pi^0_D$ events,
 and the branching ratio for $\pi^+\pi^-e^+e^-$  
 was assumed to be given by our final result.
We considered the events with  $\chi^2_D < 17$ 
 as $\pi^+ \pi^- \pi^0_D$ events,
 and removed from the sample.
The cut rejected 99.7\% of the $\pi^+ \pi^- \pi^0_D$ background,
 while retaining 92\% of the signal.
Fig.~\ref{fig-ppee-scat}(a) shows the $\Mppee$ vs $\theta^2$ 
 scatter plot of the $\pi^+\pi^-e^+e^-$ candidate events 
 after the $\chi^2_D$ cut. 
Here the solid dots represent the vacuum data and the plus 
 the He data.
Let us first consider the vacuum data.
We notice that 
 there exists a cluster of events  (15 events) inside
 our signal box defined by $\theta^2\leq$20~mrad$^2$
 and $|M_{\pi\pi ee}-\MKL|<$ 3 $\sigma_{\MKL }$.
Fig.~\ref{fig-ppee-scat}(b) is the corresponding plot obtained
 by the $\pi^+ \pi^- \pi^0_D$ MC simulation.
Note that the MC statistics in the plot is 
 about 5 times more than the vacuum data.
We also notice that there are non-negligible backgrounds
 remaining similarly in both plots.
They are more or less uniformly distributed 
  along the $\theta^2$-direction, and form two distinct bands.
It is found from the MC study that there exist two 
 mechanisms for the $\pi^+ \pi^- \pi^0_D$ events to pass
 through the $\chi^2_D$ cut.
One is $\gamma$-radiation: events in which $e^+/e^-$
 radiates $\gamma$-rays internally or externally
 would loose its energy.
Such events constitute the low mass background band.
The other is $\pi$-decay: events in which $\pi^+/\pi^-$ decays 
 inside the magnet would have wrong momentum assignment.
The high mass backgrounds are due mainly to these events.
Next we consider the He data.
There is no event left inside the signal box~\cite{note-consistency}, 
 as was reported in Ref.~\cite{Nomura97}.
The backgrounds outside the signal box are spread
 more broadly compared with the vacuum data.
These events are most probably due to the nuclear 
 interactions off the helium nucleus.

\begin{figure*}[tbp]
 \begin{center}
  \includegraphics[scale=0.8, bb=0 0 338 238, clip]{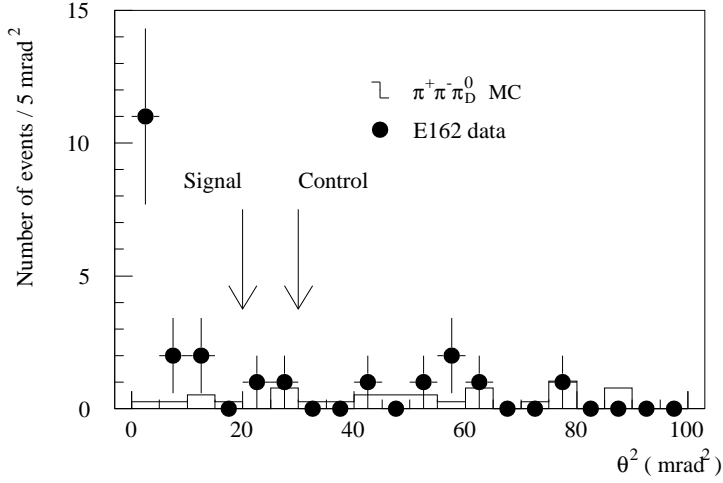}
  \caption{ The $\theta^2$ projection of 
  the $K_L\rightarrow\pi^+\pi^-e^+e^-$ candidate events.
  The solid line in the figure represents 
  the $\pi^+ \pi^- \pi^0_D$ MC simulation.
  Its $K_L$ flux was normalized by the reconstructed 
  $\pi^+ \pi^- \pi^0_D$ events.
  The arrows indicate the signal and control region boundaries.}
  \label{fig-ppee-proj}
 \end{center}
\end{figure*}

We now want to estimate the number of backgrounds inside the signal box.
To this end, we relied upon the fact that, for both He and vacuum data,
 the background distribution was approximately flat along the 
 $\theta^2$-axis.
We thus projected the events with $|\Mppee-\MKL|<3 \sigma_{\MKL }$
 onto the $\theta^2$-axis.
The resultant distribution is shown in Fig.~\ref{fig-ppee-proj}.
There is a clear signal peak at $\theta^2=0$.
The solid line in the figure represents the 
 $\pi^+ \pi^- \pi^0_D$ MC simulation, 
 in which the $K_L$ flux was normalized with the reconstructed  
 $\pi^+ \pi^- \pi^0_D$ events.
We defined a  ``control'' region to be $30<\theta^2<100$~mrad$^2$.
We then determined a scale factor by normalizing the number of 
 $\pi^+ \pi^- \pi^0_D $ MC events in this region (5.2 events) 
 to that of the data (6 events).
The number of backgrounds was given by the number of 
 $\pi^+ \pi^- \pi^0_D $ MC events 
 in the signal region (1.3 events) times the scale factor (1.15).
We found this to be $ 1.5 \pm 1.0$, where the error includes 
 uncertainty due to the MC statistics.
Subtracting this from the events inside the signal box, we
 determined the number of signal events to be $13.5 \pm 4.0 $.

\section{Branching Ratio and Systematic Errors}
The branching ratio was calculated by
\begin{eqnarray*}
Br(K_L \rightarrow \pi^+ \pi^- e^+ e^-)=
Br(K_L \rightarrow \pi^+ \pi^- \pi^0) \times
Br(\pi^0 \rightarrow e^+ e^- \gamma) \\
\times
\frac{   A(\pi^+\pi^-\pi^0_D)}{   A(\pi^+\pi^-e^+e^-)} \cdot
\frac{\eta(\pi^+\pi^-\pi^0_D)}{\eta(\pi^+\pi^-e^+e^-)} \cdot
\frac{   N(\pi^+\pi^-e^+e^-)}{   N(\pi^+\pi^-\pi^0_D)}
\end{eqnarray*}
 where $A$, $\eta$ and $N$ denote acceptance, efficiency and
 observed number of events, respectively.
The detector acceptances were determined by MC simulations, 
 and were found to be 
 $0.98 \times 10^{-3}$  for $\pi^+ \pi^- \pi^0_D$ and 
 $2.6 \times 10^{-3}$   for $\pi^+\pi^-e^+e^-$~\cite{note-accep-change}.
Here, the latter acceptance depends upon 
 the $\pi^+\pi^-e^+e^-$ matrix elements; 
 we employed a theoretical model given by Ref.~\cite{SW92}.
Most of efficiencies were common to the both modes, and they tended to
 cancel out in the efficiency ratio.
The largest difference stemmed from the detection efficiency of 
 extra $\gamma$-rays in $\pi^+ \pi^- \pi^0_D$.
Using a MC simulation, we estimated its inefficiency to be 20.4\%.
The $\chi^2_D$ cut and  $\pi^0 \rightarrow e^+ e^- \gamma$
 inclusive cut were applied only to the signal mode: 
 the efficiency of the former was found to be 92\%
 while the over-veto due to the latter was estimated to be 1\%.
We found that all other effects caused negligibly small difference in
 the ratio of efficiency.
Using the known branching ratio of 
 $ Br(K_L \rightarrow \pi^+ \pi^- \pi^0)=0.1256$ and
 $ Br(\pi^0 \rightarrow e^+ e^- \gamma)=1.198 \times 10^{-2}$~\cite{PDG},
 we finally arrived at the branching ratio of %
\[
  Br( K_L \rightarrow \pi^+ \pi^- e^+ e^-)
   = ( 4.4 \pm 1.3 \pm 0.5) \times 10^{-7} \ ,
\]
where the first (second) error represents the statistical
 (systematic) uncertainty.

Table~\ref{table-sys-error} summarizes our systematic errors, 
 which are divided into three categories.
 \begin{table}
  \caption{Summary of systematic errors in the branching ratio.}
  \label{table-sys-error}
   \begin{center}
    \begin{tabular}[htbp]{lr}
     \hline \hline
     Source & uncertainty \\ \hline \hline
     $K_L$ momentum spectrum   &   4.8\% \\
     Matrix element            &   3.9\% \\
     Others                    &   3.1\% \\ \hline
     Background subtraction    &   7.4\% \\
     Nuclear interaction       &   3.6\% \\
     Other contamination       &   1.4\% \\ \hline
     $Br(K_L \rightarrow \pi^+ \pi^- \pi^0_D)$
                               &   3.1\% \\ \hline
     Total                     &  11.3\% \\
     \hline \hline
    \end{tabular} \end{center}
   \begin{center}
   \end{center}
 \end{table}
The first one is related to the acceptance-efficiency ratio.
The largest contribution in this category came from 
 uncertainty in the $K_L$ momentum spectrum employed 
 as an input to the various MC studies.
It affected both acceptance and efficiency,
 and resulted in the fractional uncertainty of 4.8\% in the 
 final branching ratio.
The second largest contribution stemmed from uncertainty in 
 the matrix element of the theoretical model.
This brought 3.9\% uncertainty in the acceptance ratio.
The second category is related to the number of the  events
 $N(\pi^+\pi^-e^+e^-)$ and/or $N(\pi^+\pi^-\pi^0_D)$.
We checked stability of the background subtraction by 
 employing different control regions.
Actually, widening the region in the $\Mppee$ direction,
 we tried $|\Mppee-\MKL|<5 \sigma_{\MKL }$ and 
  $<7 \sigma_{\MKL }$ instead of  $<3 \sigma_{\MKL }$.
We considered the biggest difference resulted from this treatment as 
 a systematic error (7.4\%).
For the He data, 
we also tested different event shapes assumed for the backgrounds
 due to the nuclear interaction.
The effect was at most 3.6\%
 as long as they gave statistically consistent results 
 with the null observation.
All other background contaminations, 
 such as external $\gamma$-ray conversion events 
 associated with $K_L \rightarrow \pi^+ \pi^- \gamma$ and 
 $K_L \rightarrow \pi^+ \pi^- \pi^0$ 
 ($\pi^0 \rightarrow \gamma \gamma $),
 were estimated to be small (1.4\%).
The final category is related to the current experimental errors
 on the branching ratios $Br(K_L \rightarrow \pi^+ \pi^- \pi^0) $ and
 $Br(\pi^0 \rightarrow e^+ e^- \gamma)$.
Summing up all the uncertainties in quadrature,
 we found the over all systematic error to be 11.3\%.

In summary, we observed $13.5 \pm 4.0 $ 
 $K_L \rightarrow \pi^+ \pi^- e^+ e^-$ events,
 and determined its branching ratio to be
  $ (4.4 \pm 1.3(\mbox{stat.}) \pm 0.5(\mbox{syst.}) ) \times 10^{-7}$.
The result is found to agree with the theoretical prediction as
  well as the recent measurement.
\begin{ack}
We wish to thank Professors H. Sugawara, S. Yamada,  S. Iwata,
 K. Nakai, and K. Nakamura
 for their support and encouragement. 
We also acknowledge the support from the operating crew of the
 Proton Synchrotron, 
 the members of Beam Channel group, Computing Center and 
 Mechanical Engineering Center 
 at KEK. 
Y.T, Y.M and M.S acknowledge receipt of Research Fellowships
 of the Japan Society for the Promotion of Science for Young Scientists.
\end{ack}


\begin{thebibliography}{99}
\bibitem{Nomura97}
 T. Nomura {\em et al.},
 Phys. Lett. B {\bf 408}, 445 (1997)
%
\bibitem{SW92}
 L. M. Sehgal and M. Wanninger,
 Phys. Rev. D{\bf 46}, 1035 (1992); {\bf 46}, 5209(E)(1992); 
 P. Heilinger and L. M. Sehgal, {\em ibid.} {\bf 48}, 4146 (1993).
%
\bibitem{OtherTh}
 D. P. Majumdar and J. Smith, 
 Phys. Rev. {\bf 187}, 2039 (1969);
%% J. K. Elwood, M. B. Wise, and M. J. Savage,
%% Phys. Rev. D{\bf 52}, 5095 (1995);
%% J. K. Elwood, M. B. Wise, M. J. Savage, and J. W. Walden,
%% Phys. Rev. D{\bf 53}, 4078 (1996).
 J. K. Elwood {\em et al.}, Phys. Rev. D{\bf 52}, 5095 (1995);
 J. K. Elwood {\em et al.}, {\em ibid.} {\bf 53}, 4078 (1996).
%
\bibitem{Adams98}
J.Adams  {\em et al.}, Phys. Rev. Lett. {\bf 80}, 4123 (1998)
%
\bibitem{Sasao97}
 Preliminary results were presented 
 at the SLAC topical conference; see
 N. Sasao,
 Proceedings of the SLAC topical conference (Aug. 1997).
%
\bibitem{note-He-Dalitz}
Out of this, 2066 events came from the He data, 
 which is different from the one presented in Ref.~\cite{Nomura97}.
This is because some of the applied cuts, 
 in particular those related to track/vertex quality, 
 were tightened to further reduce backgrounds.
%
\bibitem{note-chi2D-change}
We have changed slightly the definition of $\chi^2_D$: the 
 third term is now quadratic in $\theta$.
The change does not alter any results,
 and makes the interpretation of $\chi^2_D$ more
 transparent statistically.
%
\bibitem{note-sigma-theta}
The standard deviation $\sigma_{\theta}$ was determined as follows.
We could define the quantity $\theta$ as 
 $\theta=\sqrt{\theta^{2}_{x^\prime}+\theta^{2}_{y^\prime}}$,
 where $\theta_{x^\prime}$ and $\theta_{y^\prime}$
 are projected angles onto some mutually orthogonal planes
 containing the target and vertex point.
It was found that $\theta_{x^\prime}$ and $\theta_{y^\prime}$
 were expressed by a Gauss distribution function with an 
 approximately equal r.m.s. deviation of 
 $\sigma_{\theta} \simeq 1.2$~mrad.
%
\bibitem{note-consistency}
We checked statistical consistency of the two data sets:
one observing 15 events (including backgrounds) and the other none.
The probability was found to be $\sim 6$\%, which
 is somewhat small but not unreasonable.
%
\bibitem{note-accep-change}
Apparent change in the acceptance values 
 from those quoted in Ref.~\cite{Nomura97} 
 is not essential and has occurred as follows. 
In the acceptance/efficiency calculations, 
 the cut imposed on  $M_{ee}$, 
 which previously had been applied in acceptance, 
 was treated in efficiency.
In addition,
 we used measured $K_{L}$ Dalitz plot parameters~\cite{PDG} 
 to calculate the $\pi^+ \pi^- \pi^0$ matrix element
 instead of phase space. 
%
\bibitem{PDG}
  Particle Data Group, C. Caso {\em et al.},
  Eur. Phys. J. C{\bf 3}, 1-794 (1998)

\end{thebibliography}
\end{document}